\UseRawInputEncoding

\documentclass[preprint]{aastex63}
\usepackage{multirow}
\submitjournal{ApJ}

\shorttitle{MIR Properties of Radio NLS1s}
\shortauthors{Fan et al.}

\begin{document}

\title{Mid-Infrared Properties of Narrow-Line Seyfert 1 Galaxies Detected by LoTSS DR2}

\correspondingauthor{Xu-Liang Fan}
\email{fanxl@sues.edu.cn}

\author[0000-0003-0988-9910]{Xu-Liang Fan}
\affiliation{School of Mathematics, Physics and Statistics, Shanghai University of Engineering Science, Shanghai 201620, China}
\affiliation{Center of Application and Research of Computational Physics, Shanghai University of Engineering Science, Shanghai 201620, China}



\begin{abstract}
  Narrow-line Seyfert 1 galaxies (NLS1s), a subclass of active galactic nuclei (AGNs) at early stage of accretion process, are also found to host relativistic jets. However, currently known jetted NLS1s are rare. The majority of NLS1s are undetected at radio band. The radio detection rate of NLS1s raises with the LOFAR Two-metre Sky Survey (LoTSS), which gives a good opportunity to find more jetted NLS1s. The better sensitivity brings another question of whether the radio emission of NLS1s with low radio luminosity originates from jet activity. In order to clarify the origin of radio emission for NLS1s, and search for more jetted NLS1s, we explore the mid-infrared properties of LoTSS detected NLS1s by comparing them with known jetted AGNs and star forming galaxies (SFGs), which locate above and on the well studied radio/far-infrared correlation, respectively. The majority of NLS1s show mid-infrared (MIR) excess compared with SFGs. Their radio emission shows significant correlation with MIR emission. In MIR color-color diagram, NLS1s are overlapped with flat spectrum radio quasars, but well separated from SFGs and optically selected radio galaxies. The flux ratio between radio and MIR emission of these NLS1s is also similar with a radio quiet quasar with weak jet. These results imply substantial contributions from AGN activities for both radio and MIR emission of NLS1s. A small fraction of NLS1s with relatively higher radio luminosity locate in similar region with blazars in radio-MIR diagram, which suggests that the radio emission of these NLS1s is dominated by jet. We obtain a sample of jetted NLS1 candidates through their radio excess in radio-MIR diagram.
\end{abstract}



\section{Introduction} \label{sec:intro}
Bright extragalactic radio sources ($\gtrsim$ 1 Jy) are believed to be produced by powerful relativistic jets in active galactic nuclei (AGNs,~\citealt{2016A&ARv..24...13P}). They typically show extended radio lobes, or extreme features due to the Doppler beaming effect\citep{2019ARA&A..57..467B}. As the sensitivity of radio observations improves, the origin of radio emission in faint extragalactic radio sources becomes complicated~\citep{2016A&ARv..24...13P}. Components related to star formation (SF) activity, disk wind and disk-coronal activity are all suggested to explain their radio emission~\citep{2019NatAs...3..387P}. Star forming galaxies (SFGs) are even suggested to dominate radio sky when the 1.4 GHz radio luminosity is fainter than $\sim 10^{30} erg~s^{-1} Hz^{-1}$~\citep{2016A&ARv..24...13P}. On the other side, low power extended jet structures were indeed found, which might break this simple luminosity criterion~\citep{2019MNRAS.488.2701M, 2021MNRAS.500.4921W}. As most radio sources are still unresolved in radio surveys~\citep{2019A&A...622A...1S, 2022A&A...659A...1S}, additional methods are needed to distinguish radio emission dominated by AGN/jet from that dominated by SF activity (e.g.\citealt{2016A&ARv..24...13P, 2022MNRAS.512..296L}).

Optical spectroscopic features are widely applied to distinguish radio AGNs from SFGs, such as  the BPT diagram~\citep{1981PASP...93....5B}, the relationship between the strength of 4000 $\AA$ break and specific radio luminosity (the ratio of radio luminosity to stellar mass,~\citealt{2005MNRAS.362....9B}), and the correlation between emission line (usually H$\alpha$) and radio luminosity~\citep{2012MNRAS.421.1569B}. Radio loudness, radio spectral properties, X-ray emission and infrared colors are also available to select AGNs~\citep{2016A&ARv..24...13P}. Apart from these methods, the radio/far-infrared (FIR) correlation is also frequently used to clarify the origin of radio emission based on the assumption that FIR emission is dominated by the dust heated by SF process (e.g., \citealt{1985A&A...147L...6D, 1985ApJ...298L...7H, 2010MNRAS.409...92J, 2015MNRAS.453.1079B, 2018MNRAS.475.3010G}). The jet induced radio emission results in obvious radio excess in radio-FIR diagram, while the radio emission from SF activity follows a significant correlation with FIR emission~\citep{2015MNRAS.453.1079B, 2016MNRAS.462.1910Hd}. This relation has been confirmed between FIR band and low or high radio frequencies, respectively~\citep{2018MNRAS.475.3010G, 2018MNRAS.480.5625R}. Due to the lack of FIR observations and the large sky coverage of Wide-field Infrared Survey Explorer (\textit{WISE}) survey~\citep{2010AJ....140.1868W}, the MIR observation of \textit{WISE} mission was also suggested to constrain the origin of radio emission~\citep{2015MNRAS.451.1795C, 2016MNRAS.462.2631M, 2021ApJ...910...64K}. At MIR band, especially the W3 (12 $\mu m$) and W4 (22 $\mu m$) bands of \textit{WISE}, SF component is also dominant for SFGs~\citep{2012ApJ...748...80D, 2013ApJ...778...94R}. Thus a correlation between radio and MIR emission is also expected for them. On the other side, the MIR bandpasses of \textit{WISE} mission contain complicated radiation components besides those related to SF process, such as starlight from stellar populations, polycyclic aromatic hydrocarbon emission band, dusty torus of AGN, as well as non-thermal radiation from jet activity~\citep{2011ApJ...740L..48M, 2012ApJ...748...80D, 2013ApJ...778...94R, 2015MNRAS.451.1795C}. Therefore, direct comparisons of radio/MIR correlation for sources with distinct radiation components are needed.


After the discovery of the $\gamma$-ray emission from narrow-line Seyfert 1 galaxies (NLS1s), they became an important population of jetted AGNs~\citep{2009ApJ...699..976A, 2009ApJ...707L.142A}. NLS1s are characterized by their narrow profiles of broad Balmer lines (with the full width at half maximum less than $2000~$km s$^{-1}$), and strong Fe{\sc ii} multiple complex (with the
ratio of the Fe{\sc ii} to the H$\beta$ fluxes, R4570 $\gtrsim$ 0.5,~\citealt{1985ApJ...297..166O, 2001A&A...372..730V}). Their central engine are believed to be powered by lighter black holes (the mean value of black hole mass $\sim 10^{6.5}M_{\sun}$) and higher accretion rates (the mean value of Eddington ratio $\sim 0.79$,~\citealt{2006ApJS..166..128Z, 2012AJ....143...83X}). Radio-loud (RL) NLS1s were found to exhibit several blazar-like features~\citep{2008ApJ...685..801Y, 2017A&A...603A.100L}, and are believed to be the low power counterparts of most powerful jetted AGNs~\citep{2015A&A...575A..13F}.~\citet{2015A&A...575A..13F} and~\citet{2015A&A...578A..28B} explored the properties of RL NLS1s and found that misaligned jetted NLS1s were missing in current NLS1 samples, according to the unification model for relativistic jet~\citep{1995PASP..107..803U}. In order to solve this problem, large sample of jetted NLS1s is needed. Despite of the extreme properties of $\gamma$-ray detected NLS1s~\citep{2019ApJ...872..169P} and the large sample of NLS1s in optical surveys~\citep{2006ApJS..166..128Z, 2017ApJS..229...39R}, most NLS1s are found to be radio quiet~\citep{2006AJ....132..531K}, even undetected at radio band~\citep{2018A&A...615A.167C, 2018MNRAS.480.1796S}. The radio detection rate appears to increase when the sensitivity of radio observations improves~\citep{2020MNRAS.498.1278C}.~\citet{2020Univ....6...45F} found that 28.1 percent of NLS1s were detected by the high sensitivity low-frequency radio observation of LOFAR Two-metre Sky Survey (LoTSS,~\citealt{2019A&A...622A...1S}). The radio luminosity of LoTSS detected NLS1s can be as low as $10^{29} erg s^{-1} Hz^{-1}$ at 144 MHz. This brings another issue of whether jets exist in these low luminosity radio NLS1s. It is necessary to clarify origin of radio emission for low luminosity radio NLS1s without resolved radio morphology~\citep{2022A&A...658A..12J}. Moreover, potential low power jet in NLS1s can shed light on the formation and growth of jet in rapidly accreting systems.

In this paper, we compare the radio/MIR correlation and the MIR color properties of LoTSS detected NLS1s with those of known SFGs and jetted AGNs, in order to clarify the origin of their radio emission, and find more jetted NLS1s candidates. In section~\ref{sec:sample}, we introduce the samples used in our analysis. Section~\ref{sec:correlation} shows the results of radio/MIR correlations, while the properties of MIR colors are shown in section~\ref{sec:color}. We discuss the implications of our results and give a sample of jetted NLS1 candidates in section~\ref{sec:discussion}. The main conclusions are summarized in section~\ref{sec:summary}. A $\Lambda$CDM cosmology model with $\Omega_{\lambda} = 0.7$, $\Omega_{m} = 0.3$ and $H_{0} = 70~km s^{-1} Mpc^{-1}$ is used in our calculations. The spectral index of power-law spectral energy distribution (SED) is defined as $F_{\nu} \propto \nu^{-\alpha}$ throughout this paper.

\section{Sample Construction} \label{sec:sample}
For NLS1s, SF process could be an important source of their radio emission, apart from relativistic jet~\citep{2015MNRAS.451.1795C, 2022A&A...658A..12J}. Therefore, we construct comparison samples with dominant SF and jet components, respectively, then compare their radio and MIR properties with a sample of NLS1s. The comparison samples include a sample of SFGs which reveals pure contribution from SF process at radio and MIR bands~\citep{2012MNRAS.421.1569B} and a sample of BL Lac objects (hereafter BZBs) whose radio and MIR emission are believed to be dominated by powerful jets~\citep{2015Ap&SS.357...75M}. A sample of flat spectrum radio quasars (BZQs) and a sample of radio galaxies (RGs), whose MIR emission might contain contributions from torus and starlight of host galaxies, respectively, are also employed to explore the influence of different origins of MIR emission on radio/MIR correlation. The samples of SFGs and RGs are drawn from~\citet[hereafter BH12 sample]{2012MNRAS.421.1569B}, which contains 2986 SFGs and 15300 narrow-line RGs, respectively. The BH12 sample was constructed by cross-matching the value-added galaxy sample from SDSS DR7 with NVSS and FIRST radio sources. SFGs and RGs were separated by careful verifications of several criteria based on optical spectroscopic information~\citep{2012MNRAS.421.1569B}. The BZBs and BZQs samples are derived from the 5th edition of \textit{Roma-BZCAT} catalogue which contains 3561 blazars (\citealt{2009A&A...495..691M, 2015Ap&SS.357...75M}, hereafter bzcat sample). Bzcat sample contains 1151 BZBs, including both BL Lac objects and BL Lac candidates, and 1909 BZQs. 11101 NLS1s selected from SDSS DR12 are used to construct NLS1 sample for our analyses~\footnote{The classification of NLS1s in R17 sample is mainly based on the FWHM of broad H$\beta$ emission line and the flux ratio of [O{\sc iii}] to H$\beta$ emission line, which relies on the accurate decomposition of various components of optical spectrum. In some cases, objects with only narrow lines or with low S/N spectra might be misclassified as NLS1s~\citep{2020CoSka..50..270B, 2024MNRAS.527.7055P}. Here we try to exclude the potentially misclassified NLS1s through a subsample of NLS1s with detected Fe{\sc ii} (R4570 $>$ 0). This subsample contains 1843 NLS1s and the results of correlation test and linear regression are consistent with those of the original samples. We retain the results of original NLS1 sample in this paper.} (\citealt{2017ApJS..229...39R}, hereafter R17 sample).

\citet{2020Univ....6...45F} explored the properties of radio detected NLS1s with LoTSS DR1 and found a much higher radio detection rate of LoTSS DR1 than that of TGSS and FIRST. LoTSS is an ongoing sky survey at 120-168 MHz which can achieve a combination of high sensitivity ($\sim$ 100 $\mu$Jy beam$^{-1}$) and high resolution (6$\arcsec$)~\citep{2019A&A...622A...1S, 2022A&A...659A...1S}. The high sensitivity of LoTSS survey gives a good opportunity to find more jetted NLS1s and explore the low power jet. Thus in this paper, the radio data from LoTSS DR2 are applied. The LoTSS DR2 catalogue (\citealt{2022A&A...659A...1S}, including 4396228 radio objects) is firstly cross-matched with AllWISE Source Catalog~\citep{2013wise.rept....1C} with the X-match in TOPCAT. The critical angular separation for cross-matching throughout this paper is determined by comparing the variation of counts of matched objects along with the separation (e.g.,~\citealt{2013ApJS..207....4M, 2018ApJ...869..133F}). Generally, the critical angular separation is chosen when the counts of matched objects show little variation. For the cross-match between LoTSS DR2 and AllWISE, the number of matched objects increases continuously until at least 15$\arcsec$. The critical angular separation of 3$\arcsec$ is chosen to take a balance between the increasing counts of matched objects and the fraction of multiple matches. The fraction of multiple matches is less than 0.1 percent when the separation is less than 3$\arcsec$. This results in the parent sample for the following sample construction (hereafter LoTSS-AllWISE sample). There are 2821904, 2764222, 1614546, and 698754 objects detected at W1, W2, W3, and W4 band in LoTSS-AllWISE sample, respectively, with S/N of AllWISE flux measurement $>$ 2 (i.e., magnitude but not upper limit is given in AllWISE\footnote{https://wise2.ipac.caltech.edu/docs/release/allwise/expsup/sec2\_1a.html}). This sample is then cross-matched with BH12, bzcat, and R17 sample with angular separation 3$\arcsec$. The numbers of multiple matches for these three samples are 44, 9 and 1, respectively, at most about one percent of whole matched sample. The samples of SFGs, RGs, BZBs, BZQs, and NLS1s obtained after cross-matching are used for our analyses. The composition of each sample is summarized in table~\ref{tab:sample}.

\begin{deluxetable*}{cccccccc}
\tablecaption{The Sample Compositions and Correlation Test Results \label{tab:sample}}
\tabletypesize{\scriptsize}
\tablehead{
\colhead{Sample} & \colhead{WISE Band} & \colhead{Source Number} & \multicolumn{2}{c}{Flux Correlation} & \colhead{Source Number with z} & \multicolumn{2}{c}{Luminosity Partial Correlation} \\
\colhead{} & \colhead{} & \colhead{} & \colhead{$r$} & \colhead{P} & \colhead{} & \colhead{$r$} & \colhead{P}
}
\startdata
\multirow{6}{*}{NLS1s} & W1 & 1874 & 0.37 & 2.7e-61 & 1874 & 0.34 & 4.6e-52 \\
 & W2 & 1874 & 0.35 & 8.5e-56 & 1874 & 0.31 & 1.2e-43 \\
 & W3 & 1862 & 0.41 & 3.4e-76 & 1862 & 0.38 & 3.6e-65 \\
 & W4 & 1686 & 0.47 & 6.2e-92 & 1686 & 0.45 & 2.5e-85 \\
 & W1+W2+W3 & 1862 & \nodata & \nodata & \nodata & \nodata & \nodata \\
 & W1+W2+W3+W4 & 1684 & \nodata & \nodata & \nodata & \nodata & \nodata \\
\hline
\multirow{6}{*}{SFGs} & W1 & 1057 & 0.43 & 3.6e-49 & 1057 & 0.40 & 1.4e-42 \\
 & W2 & 1057 & 0.49 & 6.6e-64 & 1057 & 0.45 & 5.3e-54 \\
 & W3 & 1053 & 0.56 & 1.4e-87 & 1053 & 0.53 & 8.3e-76 \\
 & W4 & 1035 & 0.50 & 1.2e-65 & 1035 & 0.45 & 1.2e-53  \\
 & W1+W2+W3 & 1053 & \nodata & \nodata & \nodata & \nodata & \nodata \\
 & W1+W2+W3+W4 & 1035 & \nodata & \nodata & \nodata & \nodata & \nodata \\
\hline
\multirow{6}{*}{RGs} & W1 & 4266 & 0.02 & 0.23 & 4266 & 0.19 & 7.1e-37 \\
 & W2 & 4267 & 0.03 & 0.05 & 4267 & 0.20 & 4.1e-40 \\
 & W3 & 1894 & 0.007 & 0.77 & 1894 & 0.07 & 0.001 \\
 & W4 & 478 & 0.08 & 0.07 & 478 & 0.12 & 0.008 \\
 & W1+W2+W3 & 1893 & \nodata & \nodata & \nodata & \nodata & \nodata \\
 & W1+W2+W3+W4 & 438 & \nodata & \nodata & \nodata & \nodata & \nodata \\
\hline
\multirow{6}{*}{BZBs} & W1 & 282 & 0.55 & 1.5e-23 & 120 & 0.65 & 9.6e-16 \\
 & W2 & 282 & 0.58 & 3.5e-27 & 120 & 0.66 & 4.3e-16 \\
 & W3 & 248 & 0.55 & 4.9e-21 & 97 & 0.61 & 5.6e-11 \\
 & W4 & 162 & 0.46 & 1.0e-9 & 54 & 0.62 & 7.3e-7 \\
 & W1+W2+W3 & 248 & \nodata & \nodata & \nodata & \nodata & \nodata \\
 & W1+W2+W3+W4 & 162 & \nodata & \nodata & \nodata & \nodata & \nodata \\
\hline
\multirow{6}{*}{BZQs} & W1 & 393 & 0.16 & 0.001 & 393 & 0.15 & 0.004 \\
 & W2 & 393 & 0.19 & 2.2e-4 & 393 & 0.18 & 4.3e-4 \\
 & W3 & 386 & 0.20 & 1.1e-4 & 386 & 0.18 & 2.8e-4 \\
 & W4 & 321 & 0.21 & 1.7e-4 & 321 & 0.20 & 3.8e-4 \\
 & W1+W2+W3 & 386 & \nodata & \nodata & \nodata & \nodata & \nodata \\
 & W1+W2+W3+W4 & 320 & \nodata & \nodata & \nodata & \nodata & \nodata \\
\enddata
\end{deluxetable*}

\section{Correlations between radio and MIR emission} \label{sec:correlation}
\begin{figure*}[h!]
\centering
\includegraphics[angle=0,scale=0.55]{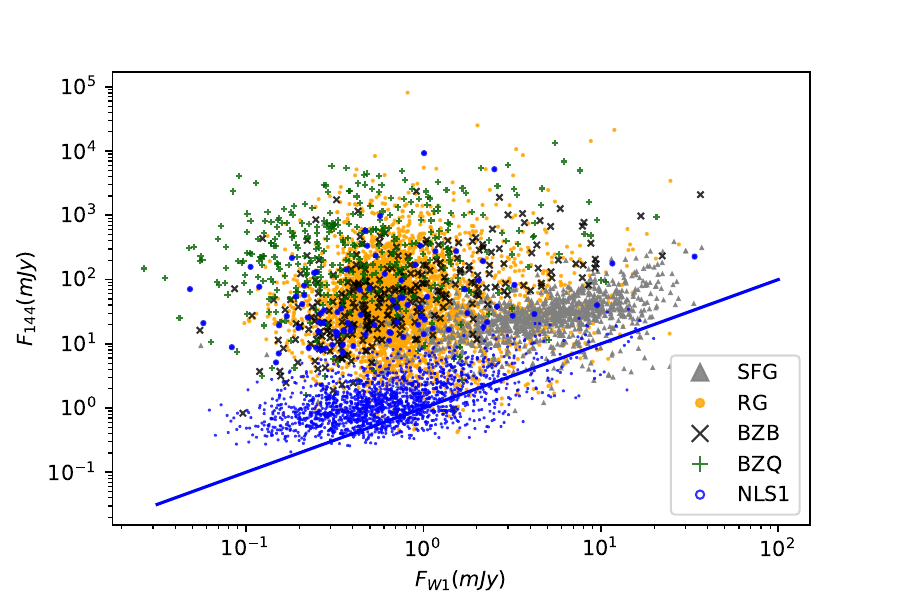}
\includegraphics[angle=0,scale=0.55]{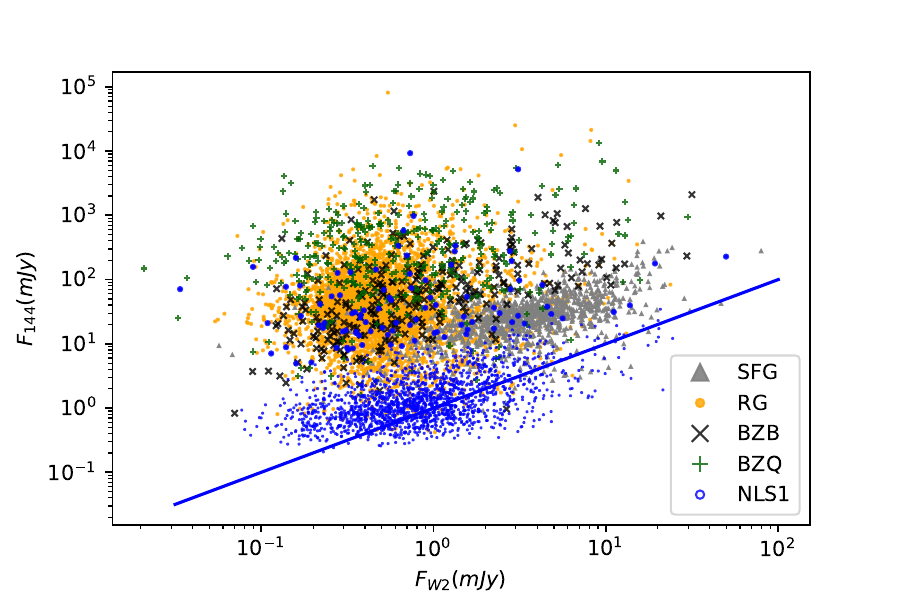}
\includegraphics[angle=0,scale=0.55]{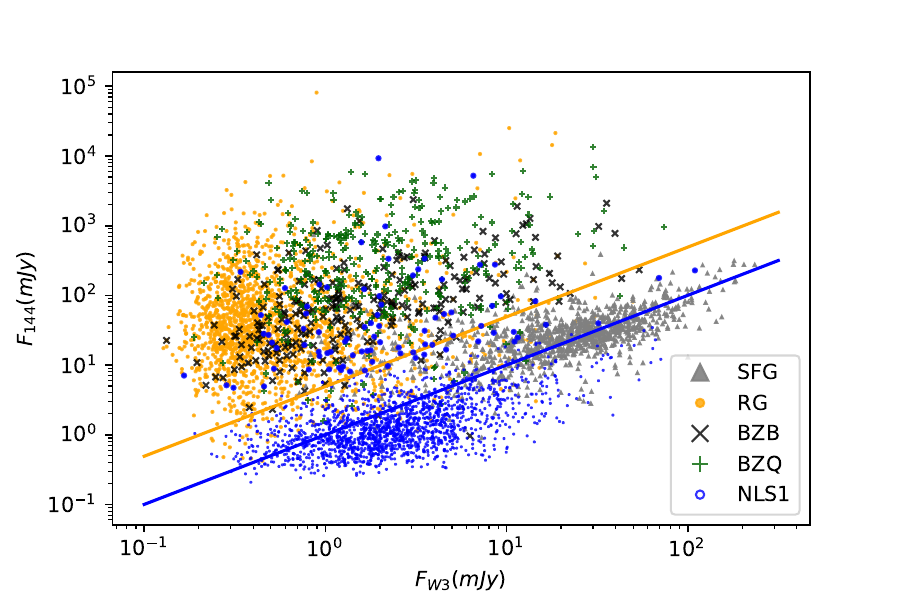}
\includegraphics[angle=0,scale=0.55]{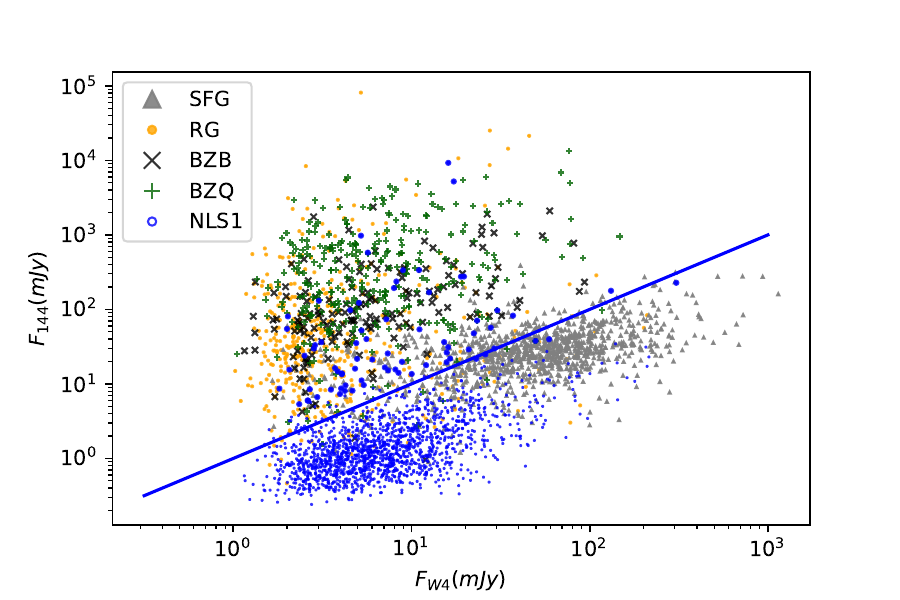}
\caption{144 MHz radio flux density versus MIR flux density at W1, W2, W3, W4 bands, respectively. The blue open circles in each panel represent NLS1s, while the blue filled circles with larger symbol sizes label the outliers. The grey triangles represent SFGs. The orange filled circles represent RGs, the black oblique crosses represent BZBs and the green crosses represent BZQs. The blue solid line in each panel represents the one to one relation. The orange solid line in the bottom left panel represents that 1.4 GHz radio flux density (which is converted into 144 MHz flux density with $\alpha_{radio} = 0.7$) equals to W3 flux density. \label{fig:flux}}
\end{figure*}

SFGs are expected to show significant correlation between radio and MIR emission due to their common origin from SF process. We firstly explore the radio/MIR correlation for SFGs with 144 MHz radio flux density and MIR flux density at W1 (3.4 $\mu m$), W2 (4.6 $\mu m$), W3 (12 $\mu m$) and W4 (22 $\mu m$) bands, respectively, where the Vega magnitude in AllWISE catalog is converted to flux density with zero magnitude flux density 309.540 Jy, 171.787 Jy, 31.674 Jy, 8.363 Jy for W1, W2, W3, and W4 bands, respectively~\citep{2011ApJ...735..112J}. The scatters are plotted in Figure~\ref{fig:flux}. The blue solid line in each panel shows that radio flux density equals to MIR flux density. The MIR flux density of SFGs increases from short to long wavelength obviously. The 144 MHz radio flux density is generally larger than the MIR flux density at W1 and W2 bands. The $12~\mu m$ MIR flux density is comparable with 144 MHz radio flux density (with the mean values of $F_{12\mu m}$ and $F_{144 MHz}$ equalling to 19.01 mJy and 24.71 mJy, respectively), while the $22~\mu m$ flux density is generally larger than radio flux density (43.75 mJy versus 24.97 mJy). The increasing flux density from W1 to W4 band results in slight sources number reduction due to the poor sensitivity of W4 band (1035/1057 for W4/W1, respectively,~\citealt{2010AJ....140.1868W}). As the radio and IR emission are found to be linearly correlating with star formation rate (SFR) in logarithmic space~\citep{2015MNRAS.453.1079B, 2019A&A...631A.109W}, we perform Pearson linear correlation tests to examine the correlation between radio and MIR emission for SFGs. The results of Pearson linear correlation test are listed in Table~\ref{tab:sample}. MIR Flux density of all four \textit{WISE} bands show significant correlations with radio flux density. 12 $\mu m$ MIR flux density shows the highest correlation coefficient with 144 MHz radio flux density.

RGs, BZBs and BZQs locate in the similar region in radio-MIR flux diagram, with radio flux density obviously exceeding MIR flux density for all four \textit{WISE} bands. About 82 percent (321/393) of BZQs detected at W1 band are detected at W4 band, while this proportion reduces to 57 percent for BZBs (162/282) and 11 percent (478/4266) for RGs. The linear correlation tests show significant correlations for BZBs, which show little dependence on MIR band. Weak correlations are shown for BZQs, but no correlation is found for RGs.

NLS1s show two populations in radio-MIR flux diagram. The majority of NLS1s show similar trends with SFGs that MIR flux density of long wavelength is larger than that of short ones. And 90 percent of sources are detected at W4 band compared with the sample detected at W1 band. The ratio of MIR flux density to radio flux density of NLS1s is larger than that of SFGs, e.g., for W3 band, the majority of NLS1s have 144 MHz radio flux density lower than 12 $\mu m$ MIR flux density (1.54 mJy versus 2.40 mJy), while SFGs distribute around $F_{144 MHz} =  F_{12 \mu m}$. This means NLS1s have MIR excess compared with SFGs. The remaining small fraction of NLS1s show similar radio excess with blazars and RGs. The Pearson linear correlation test between radio and MIR flux density is performed firstly for the whole NLS1 sample. The correlation coefficients generally increase from short to long wavelength, although they are all lower than those of SFGs.

The two populations of NLS1s in radio-MIR flux diagram indicate two distinct radiation components of them. In order to compare the radio/MIR correlation between NLS1s and SFGs better, we apply a sigma clipping method to exclude the outliers of NLS1s, i.e., NLS1s with radio excess, and perform the correlation tests for the remaining sources (hereafter main NLS1 sample). With $\sigma = 3$, the outlier NLS1s are all objects with higher radio flux density for all four \textit{WISE} bands (shown by blue filled circles in Figure~\ref{fig:flux}). After excluding 3-$\sigma$ outliers, NLS1s show tighter correlations between radio and MIR flux densities, with the correlation coefficient of W4 band slightly larger than that of other bands (Table~\ref{tab:cor_nls1}).

\begin{deluxetable*}{cccccccc}
\tablecaption{The Correlation Tests Results of NLS1s excluding 3-$\sigma$ Outliers \label{tab:cor_nls1}}
\tabletypesize{\scriptsize}
\tablehead{
\colhead{WISE Band} & \colhead{Source Number} & \multicolumn{2}{c}{Flux Correlation} & \colhead{Source Number} & \multicolumn{2}{c}{Luminosity Partial Correlation} \\
\colhead{} & \colhead{} & \colhead{$r$} & \colhead{P} & \colhead{} & \colhead{$r$} & \colhead{P}
}
\startdata
 W1 & 1784 & 0.55 & 3.1e-139 & 1798 & 0.51 & 2.3e-119 \\
 W2 & 1782 & 0.53 & 3.1e-128 & 1805 & 0.48 & 3.8e-106 \\
 W3 & 1777 & 0.58 & 2.0e-160 & 1795 & 0.54 & 1.3e-137 \\
 W4 & 1613 & 0.59 & 2.2e-149 & 1630 & 0.56 & 1.9e-136 \\
\enddata
\end{deluxetable*}

\begin{figure*}[h!]
\centering
\includegraphics[angle=0,scale=0.55]{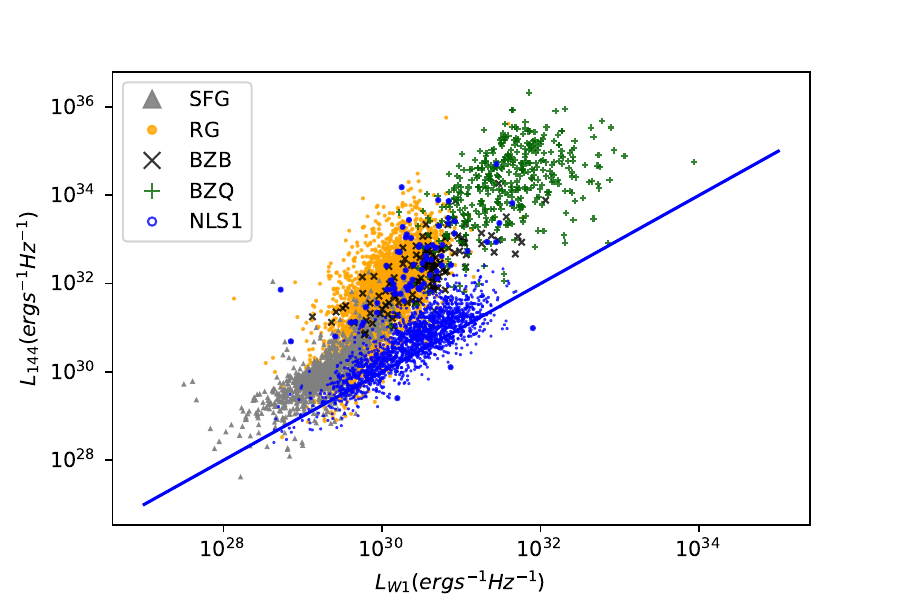}
\includegraphics[angle=0,scale=0.55]{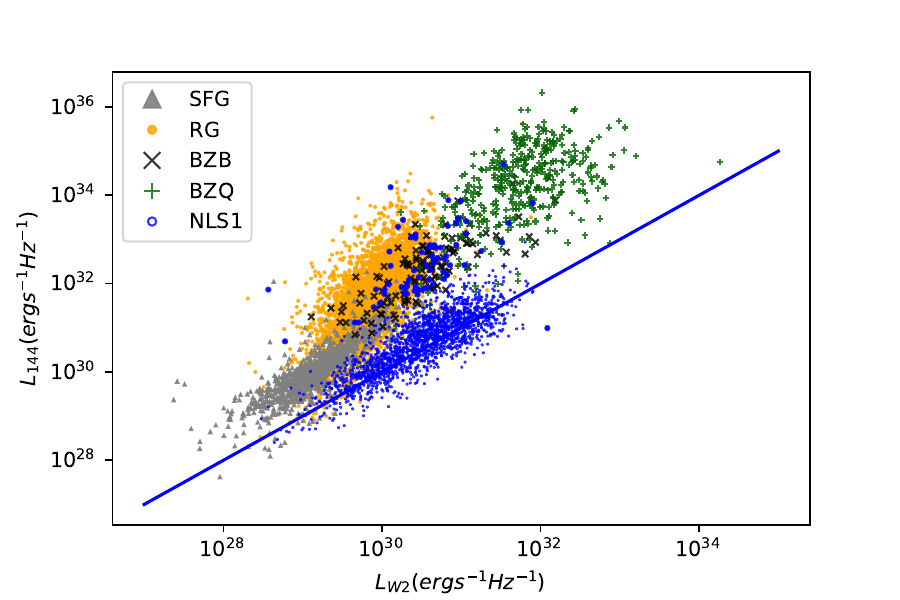}
\includegraphics[angle=0,scale=0.55]{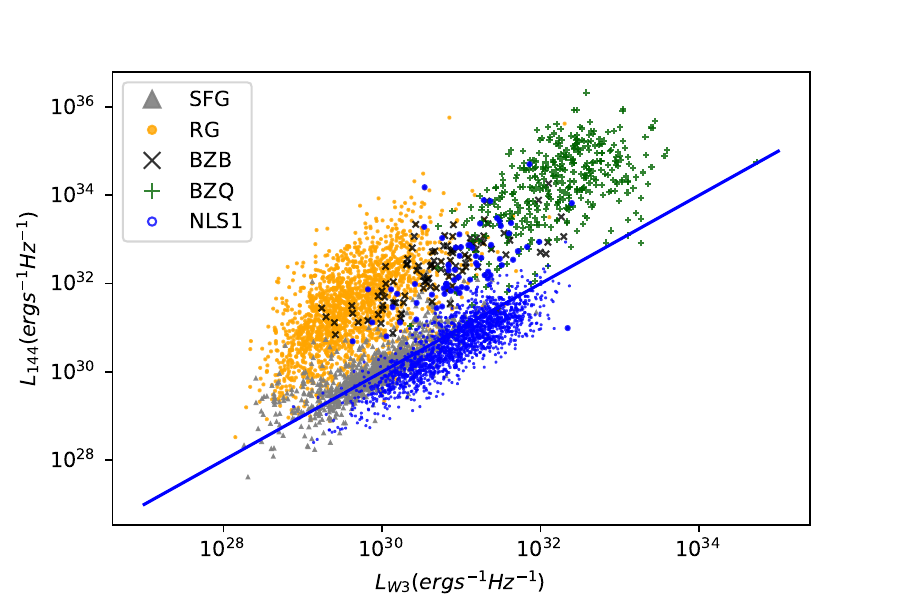}
\includegraphics[angle=0,scale=0.55]{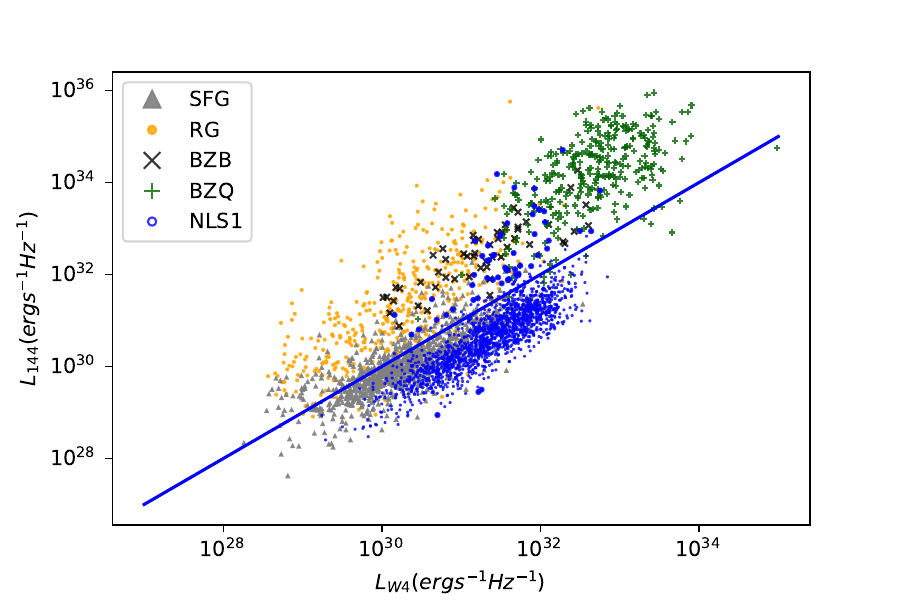}
\caption{The 144 MHz radio luminosity versus MIR luminosity at W1, W2, W3 and W4 bands, respectively. The symbols are same as Figure~\ref{fig:flux}. The blue solid lines represent the one to one relation. \label{fig:lum}}
\end{figure*}

The luminosity distributions are then compared for the five classes of sources with redsift measurements. The redshift cut only reduces the sample size of BZBs, which is also summarized in Table~\ref{tab:sample}. Motivated by the flux correlation between radio and MIR bands for SFGs and NLS1s, the monochromatic MIR luminosity with the k-correction under the assumption of a local power-law SED, rather than the SED fitted or integral IR luminosity~\citep{2012ApJ...748...80D, 2015MNRAS.453.1079B} is used. Typical power-law spectral indexes $\alpha_{IR} = 1.0$~\citep{2010AJ....140.1868W, 2011ApJ...740L..48M} and $\alpha_{radio} = 0.7$\footnote{Other spectral indexes ($\alpha_{radio}$ = 0, and $\alpha_{IR} = -2, -1, 0, 2$) are also attempted and the conclusions are not changed.}~\citep{2017A&A...598A..78I, 2022A&A...659A...1S} are used for k-correction of MIR and radio luminosity, respectively. The distribution of radio luminosity along MIR luminosity is plotted in Figure~\ref{fig:lum}. A blue solid line labelled radio luminosity equalling to MIR luminosity on each panel is also plotted. Blazars and RGs show obvious radio excess again in luminosity-luminosity diagram. The 3.4 $\mu m$ and 4.6 $\mu m$ luminosities of RGs are generally larger than those of SFGs, while their MIR luminosities are overlapped at longer wavelength. NLS1s show two distinct populations in luminosity-luminosity diagram again. The main NLS1 sample show higher radio and MIR luminosity compared with SFGs. However, they are hard to distinguish from SFGs, especially at longer wavelength. There are still dozens of NLS1s, which show radio excess and locate within the region of blazars.

In order to exclude the common dependence of radio and MIR luminosity on redshift, we apply a partial Pearson correlation test in logarithmic space to examine the correlation between radio and MIR luminosity\footnote{Partial Spearman correlation tests show similar results. As the radio and MIR luminosity is expected to be linearly correlated with SFR in logarithmic space, we retain the linear correlation test throughout this paper.}. The results for SFGs, RGs, BZBs, BZQs, and whole NLS1 sample are summarized in Table~\ref{tab:sample}, while the results for the main NLS1 sample are listed in Table~\ref{tab:cor_nls1}. BZBs show the best luminosity correlation. Radio luminosity of BZQs shows weak correlation with MIR luminosity except for W1 band. For SFGs and NLS1s, the correlation coefficients of longer wavelength are slightly higher than those of short wavelength. RGs show weak correlation at W1 and W2 bands. Note that no correlation is found between radio and MIR flux density for RGs. This is because their MIR flux density is negatively correlated with redshift, especially for W1 and W2 bands. When the influence of redshift is excluded in the partial correlation test, the correlations between radio and MIR emission at W1 and W2 bands appear.

\begin{figure*}[h!]
\centering
\includegraphics[angle=0,scale=.8]{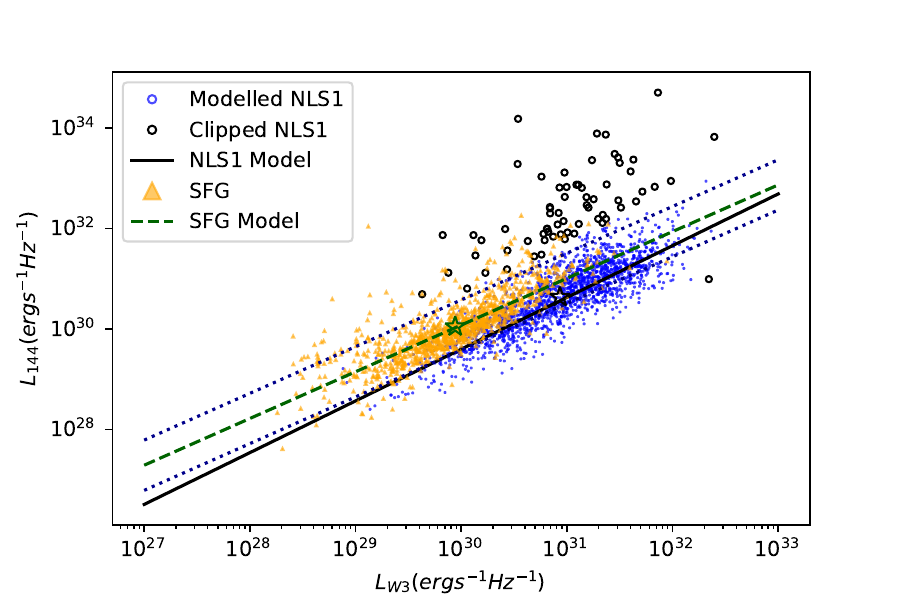}
\caption{The linear fit for NLS1s and SFGs between 144 MHz radio luminosity and 12 $\mu m$ MIR luminosity. The blue circles represent NLS1s which are fitted by linear regression, while the black circles represent the 3-$\sigma$ outliers of NLS1s. SFGs are shown with orange triangles. The black solid line shows the best fit of NLS1s, while the green dashed line shows the best fit of SFGs. The dark blue dotted lines label the 0.5 dex boundary above and below the best fit of SFGs. The green and black open stars show the centroids of SFGs and modelled NLS1s, respectively. \label{fig:lum_fit}}
\end{figure*}

For SFGs, 144 MHz radio flux/luminosity generally equals to 12 $\mu m$ MIR flux/luminosity, and the radio/MIR correlation is strongest at W3 band. Both results indicate that MIR emission at W3 band of SFGs could be the best indicator of SFR. Therefore, we perform a linear regression\footnote{https://github.com/jmeyers314/linmix}~\citep{2007ApJ...657..116K} for SFGs between 144 MHz radio luminosity and 12 $\mu m$ MIR luminosity. This gives a relation(green dashed line in Figure~\ref{fig:lum_fit})
\begin{equation}\label{eq:sfg}
\log~L_{144MHz} = (0.93 \pm 0.02) \log L_{12\mu m} + (2.18 \pm 0.60),
\end{equation}
with the intrinsic scatter 0.37 dex.

For NLS1s, We also fit the linear relation between 144 MHz radio luminosity and 12 $\mu m$ MIR luminosity after excluding the 3-$\sigma$ outliers. This results in a linear relation with slope close to unity (black solid line in Figure~\ref{fig:lum_fit}),
\begin{equation}
\log~L_{144MHz} = (1.03 \pm 0.01) \log L_{12\mu m} - (1.30 \pm 0.46).
\end{equation}
This relation has the intrinsic scatter 0.33 dex, similar with the relation of SFGs.

In Figure~\ref{fig:lum_fit}, the blue dotted lines show the additional 0.5 dex above and below the best fit of SFGs, respectively. This bound is chosen to be slightly larger than the intrinsic scatter of the relation for SFGs (0.37 dex), which is also used to separate jetted NLS1 candidates in Section~\ref{sec:discussion}. The black and green stars show the centroids of the main NLS1 sample ($\log L_{12 \mu m} = 30.93~erg s^{-1} Hz^{-1}$ and $\log L_{144 MHz} = 30.64~erg s^{-1} Hz^{-1}$) and SFGs ($\log L_{12 \mu m} = 29.94~erg s^{-1} Hz^{-1}$ and $\log L_{144 MHz} = 30.04~erg s^{-1} Hz^{-1}$), respectively. The centroid of NLS1s doses not follow the best fit of SFGs and is close to the 0.5 dex line of MIR excess side. Although SFGs and NLS1s can not distinguish from each other in the luminosity-luminosity diagram, the MIR excess of NLS1s compared with SFGs is proved\footnote{Note that NLS1s can also be considered as radio weak compared with SFGs. However, SF process can contribute both radio and MIR emission, and no evidence manifests that NLS1s is lacking SF. On the contrary, high SFR and weak jet are both reported in individual NLS1\citep{2014MNRAS.441..172C, 2021MNRAS.508.1305Y}. On the other side, torus can also contribute the MIR emission of NLS1s (Section~\ref{sec:discussion}). Thus we conclude MIR excess of NLS1s is proved.}.

\section{Properties of MIR color}\label{sec:color}
The color-color diagram among three or four \textit{WISE} bands has been widely suggested to distinguish different MIR emitters. Different extragalactic objects are expected to occupy different regions in the \textit{WISE} color-color diagram~\citep{2010AJ....140.1868W}. Elliptical galaxies with old stellar populations show both blue W1 $-$ W2 and W2 $-$ W3 colors. The W2 $-$ W3 color shifts to redder side due to the increasing SF activity from elliptical galaxies to spiral galaxies and SFGs~\citep{2011ApJ...735..112J}. AGNs generally have redder W1$-$ W2 than galaxies due to their power-law IR emission.~\citet{2012ApJ...753...30S} proposed that W1 $-$ W2 $> 0.8$ was a good indicator to select AGNs. In W1 $-$ W2/W2 $-$ W3 color diagram, blazars were found to occupy a specific region which was named \textit{WISE} blazar strip~\citep{2011ApJ...740L..48M}. This strip is consistent with the prediction of non-thermal radiation from jet, although a fraction of blazars also lie in the AGN region in the \textit{WISE} color-color diagram~\citep{2011ApJ...735..112J, 2011ApJ...740L..48M}.

\begin{figure*}[h!]
\centering
\includegraphics[angle=0,scale=.55]{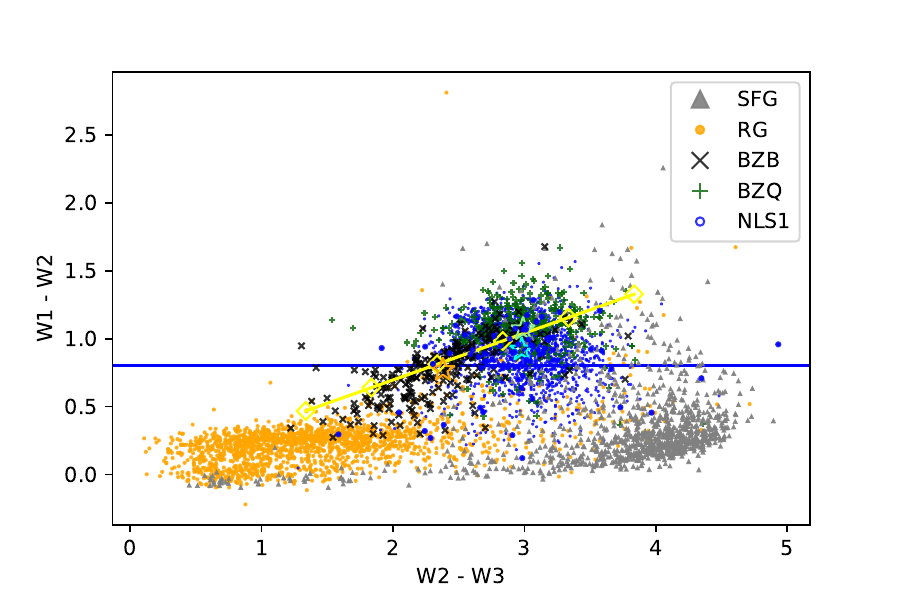}
\includegraphics[angle=0,scale=.55]{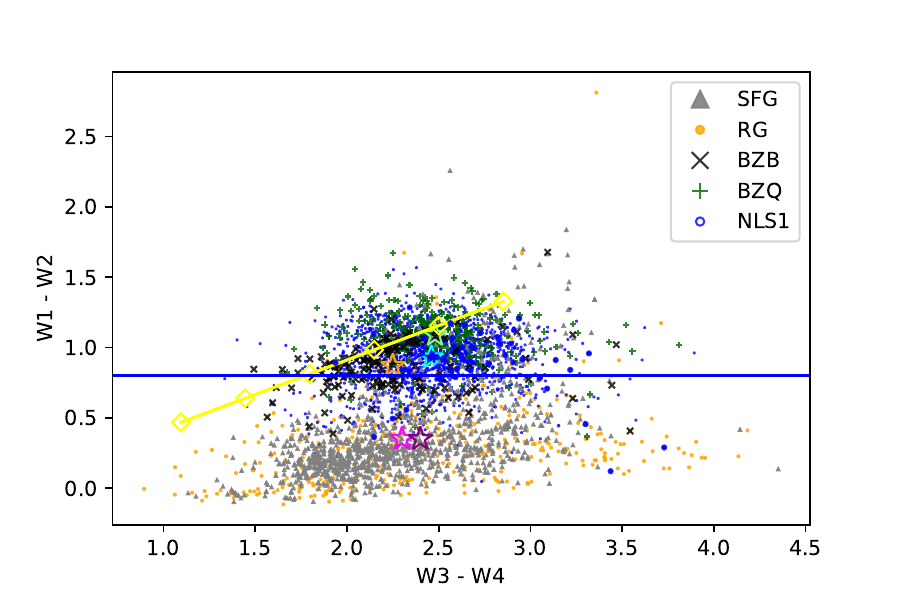}
\caption{The \textit{WISE} color-color diagram. Left panel shows W1 $-$ W2 versus W2 $-$ W3. Right panel shows W1 $-$ W2 versus W3 $-$ W4. The symbols are same as Figure~\ref{fig:flux}. The outliers of NLS1s (filled blue circles) are consistent with those of Figure~\ref{fig:lum_fit}, while the number of outliers reduces to 49 when calculating W3 $-$ W4. The blue solid lines show W1 $-$ W2  equals to 0.8 which is suggested by~\citet{2012ApJ...753...30S} as the criterion to select AGNs. The yellow line connecting open diamonds show the predicted MIR color of power-law MIR spectrum with the power-law index ranging from -0.5 to 2.0 (with step of 0.5). The cyan, light green, orange, magenta and purple stars on individual panel show the centroids of NLS1s, BZQs, BZBs, RGs and SFGs, respectively. \label{fig:color}}
\end{figure*}

The left panel of Figure~\ref{fig:color} shows the distributions of W1 $-$ W2 versus W2 $-$ W3 for the the five samples in this work. SFGs and RGs are clearly separated through W2 $-$ W3. SFGs show obvious redder W2 $-$ W3 colors than RGs, while their W1 $-$ W2 colors are both bluer than those of NLS1s and BZQs. NLS1s and BZQs generally locate at similar region in this color-color diagram. The W1 $-$ W2 colors of the majority of NLS1s are redder than the criterion to select AGNs defined by~\citet{2012ApJ...753...30S} (blue solid line in Figure~\ref{fig:color}). There are also a fraction of NLS1s showing bluer W1 $-$ W2, with W2 $-$ W3 in the middle of RGs and SFGs. The centroids of NLS1s (cyan star) and BZQs (light green star) are also plotted in Figure~\ref{fig:color}. The mean values of W2 $-$ W3 are almost the same for NLS1s and BZQs (2.98 versus 2.99), while the W1 $-$ W2 color of BZQs is slightly redder than that of NLS1s (1.07 versus 0.92). The MIR colors of BZBs extend from redder W1 $-$ W2 and W2 $-$ W3 colors to bluer ones, which generally follows the predictions of power-law MIR spectrum (represented by yellow line connecting open diamonds in the left panel of Figure~\ref{fig:color}, with the spectral index from -0.5 to 2.0).

The right panel of Figure~\ref{fig:color} shows the color-color diagram of W1 $-$ W2 versus W3 $-$ W4. NLS1s and BZQs also concentrate in the same region. RGs show similar W3 $-$ W4 colors with SFGs (magenta and purple stars, respectively). BZBs also show a strip-like distribution in the plot of W1 $-$ W2 versus W3 $-$ W4, although there are more objects deviating from the predicted line of power-law spectrum (yellow line connected open diamond in the right panel of Figure~\ref{fig:color}).

\begin{figure*}[h!]
\centering
\includegraphics[angle=0,scale=.9]{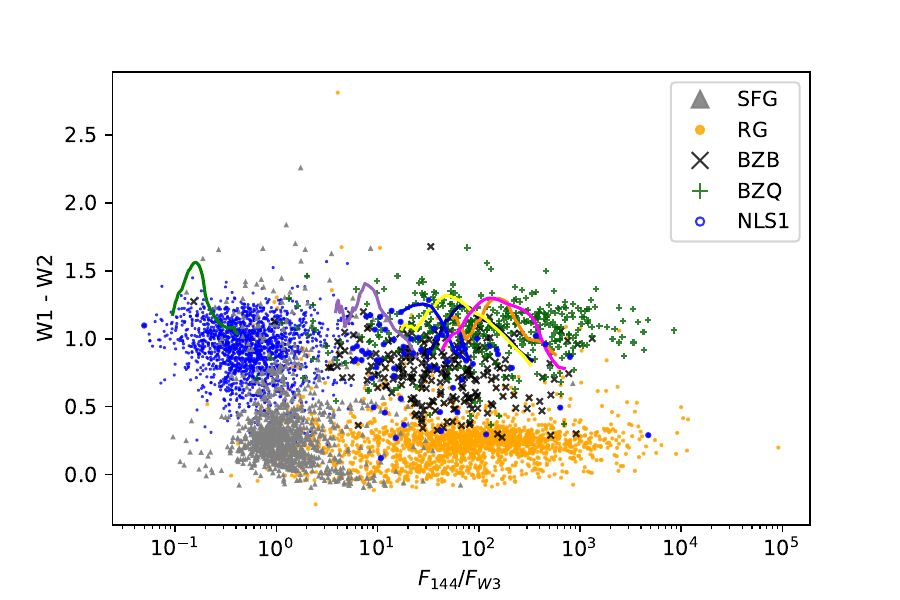}
\includegraphics[angle=0,scale=.9]{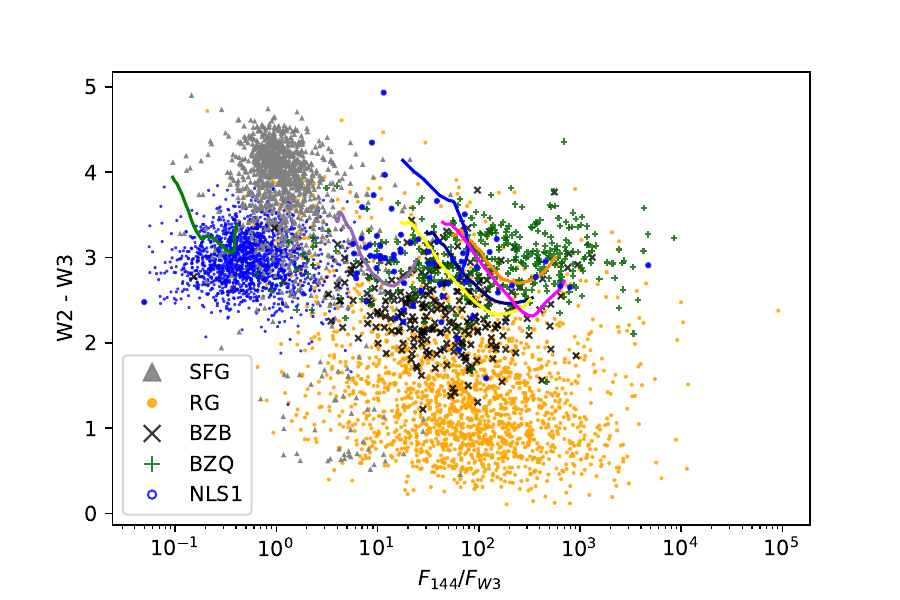}
\caption{The \textit{WISE} color versus the ratio of radio to MIR flux density. Top panel shows W1 $-$ W2 versus $F_{144MHz}/F_{12 \mu m}$. Bottom panel shows W2 $-$ W3 versus $F_{144MHz}/F_{12 \mu m}$. The symbols are same as Figure~\ref{fig:flux}. The outliers of NLS1s (filled blue circles) are consistent with those of Figure~\ref{fig:lum_fit}. The blue, dark blue, purple, orange, magenta, yellow, and green curves are derived from the SEDs of PKS 1345+12, 3C273, 3C 120, 3C 390.3, 3C 351, PG 2349-014, and Mrk 231 by varying redshifts from 0.0 to 3.0, respectively (see the text for details). \label{fig:ratio}}
\end{figure*}

In Figure~\ref{fig:ratio}, we compare the MIR color along with the ratio between 144 MHz radio flux density and 12 $\mu m$ MIR flux density. Five classes of sources in our work are generally separated in this plot. The main NLS1 sample shows slightly smaller flux ratio than SFGs (with the mean values of 0.65 and 1.29, respectively). BZQs, BZBs, and RGs have large flux ratios, while their MIR colors of both W1 $-$ W2 and W2 $-$ W3 get redder from RGs, BZBs to BZQs.

We also employ a SED library to explore the MIR color and flux ratio behaviour of different classes of sources. The Spitzer-Space-Telescope Wide-area Infrared Extragalactic Survey (SWIRE) library~\citep{2007ApJ...663...81P}, which was widely used to explore the properties of MIR color~\citep{2011ApJ...735..112J, 2015MNRAS.451.1795C}, has no wavelength coverage at radio band. Instead we use a SED library of AGNs from~\citet{2019MNRAS.489.3351B}\footnote{https://archive.stsci.edu/hlsp/agnsedatlas}. There are seven objects with wavelength coverage of radio SED longer than two meters. These seven objects include a BZQ (3C 273, e.g.,~\citealt{2017A&A...601A..35P}), two broad line RGs (3C 120, e.g.~\citealt{2017ApJ...849..146G} and 3C 390.3,~\citealt{2010A&A...509A...6B}), two steep spectrum radio quasars (SSRQs, 3C 351,~\citealt{2019MNRAS.484..385V, 2023MNRAS.519.2773B}, and PG 2349-014,~\citealt{2011A&A...534A..59G, 2011AJ....141...85R}), a gigahertz peaked-spectrum source (PKS 1345+12,~\citealt{1998PASP..110..493O}), and a radio quiet (RQ) quasar with weak jet (Mrk 231,~\citealt{2021MNRAS.504.3823W, 2021MNRAS.507.2550S}). The MIR SEDs of~\citet{2019MNRAS.489.3351B} library are based on multi-wavelength spectroscopy and photometry data, while the radio SEDs are based on several empirical models (e.g., power-law and polynomial) of discrete radio measurements. The MIR magnitude and flux density considering \textit{WISE} filter response curves are calculated with \textit{speclite} Python package\footnote{https://github.com/desihub/speclite}~\citep{2023zndo...7734526K}. Radio flux density of nearest wavelength compared with 144 MHz in the SED library is used to calculate the flux ratio\footnote{The corresponding observed wavelengths used in the calculations for different redshifts  are in range from 2 m to 2.2 m, the corresponding frequency range is from 150 MHz to 136 MHz. We also attempt to extrapolate the flux density to 144 MHz with a spectral index of 0.7, the variation of curves in Figure~\ref{fig:ratio} is very small.}. Redshift varies from 0.0 to 3.0. The results are overplotted as colored curves in Figure~\ref{fig:ratio}.

The curves of six RL AGNs are generally overlapped with BZQs, while 3C 120 has lowest flux ratio. Interestingly, the RQ quasar, Mrk 231 with detected pc- and kpc-scale jet structures~\citep{2021MNRAS.504.3823W, 2021MNRAS.507.2550S} shows flux ratio in the range of the main NLS1 sample, with slightly lower flux ratio and redder MIR colors when comparing with NLS1s.

\section{Discussion}\label{sec:discussion}
The radio/FIR correlation is widely accepted for SFGs, although potential evolutions with redshift or stellar mass are also proposed for this relation~\citep{2018MNRAS.480.5625R, 2021A&A...647A.123D}. Several works also extended this correlation to MIR band~\citep{2004ApJS..154..147A, 2013ApJ...778...94R, 2016MNRAS.462.2631M}. The tight correlations between radio and MIR emission for SFGs are confirmed in our work. This means that radio/FIR correlation can indeed extend to MIR and low-frequency radio bands. This connection is natural because SF activity can produce both radio and MIR emission. The correlation between 1.4 GHz radio luminosity and 12 $\mu m$ MIR luminosity was also found for an X-ray detected sample of SFGs by~\citet{2016MNRAS.462.2631M}. They obtained the linear relation $\log L_{1.4GHz} (erg s^{-1}) = 0.86 \log L_{12\mu m} (erg s^{-1}) + 1.4$. Although with a slightly flatter slope, their relation is generally consistent with our Equation~\ref{eq:sfg} considering a power-law index $\alpha_{radio} = 0.7$ ($\log L_{144 MHz} (erg s^{-1} Hz^{-1}) = 0.86 \log L_{12\mu m} (erg s^{-1} Hz^{-1}) + 4.47$). \citet{2022ApJ...939...26Y} derived a relation between 1.4 GHz radio luminosity and the SFR estimated from the 12 $\mu m$ MIR luminosity with the relation from~\citet{2017ApJ...850...68C}. Combining the two relations of $\log SFR (M_{\sun} yr^{-1}) = 0.8 \log L_{1.4GHz} (W Hz^{-1}) - 17.02$ and $\log SFR (M_{\sun} yr^{-1}) = 0.889 \log L_{12\mu m} (L_{\sun}) -7.76$, we get $\log L_{144 MHz} (erg s^{-1} Hz^{-1}) = 1.11 \log L_{12\mu m} (erg s^{-1} Hz^{-1}) - 3.17$. The slope is steeper than our Equation~\ref{eq:sfg}. The different slopes for different samples of SFGs might indicate evolution with redshift or SFR of radio/MIR correlation~\citep{2017ApJ...850...68C}, similar with radio/FIR correlation. In \textit{WISE} color-color diagram, SFGs concentrate in the region of starburst galaxies and luminous infrared galaxies, with W1 $-$ W2 for most of them smaller than 0.8, the criterion suggested by~\citet{2012ApJ...753...30S} to select AGNs (Figure~\ref{fig:color}).

For AGNs, radio and IR emission components become complicated. AGN activity and SF activity from host galaxy can both contribute to their radio and IR emission. A fraction of AGNs are found to show obvious radio excess beyond the radio/FIR correlation, which provides an important criterion to select jetted AGNs~\citep{2016A&ARv..24...13P}. The RGs in our work show obvious radio excess, which is consistent with the jet origin of their radio emission. However, the blue MIR color (W1 $-$ W2 and W2 $-$ W3) suggests that their MIR emission is dominated by the host starlight rather than AGN activity, and their host galaxies lack ongoing SF process.

The multi-wavelength emission of blazars is widely accepted to be dominated by non-thermal emission from jet, although the detailed radiation mechanisms and emission regions are still under debate~\citep{2019NewAR..8701541H}. The radio excess of blazars in radio-MIR diagram confirms the jet origin of their radio emission. The common radio and MIR emission from jet implies a tight correlation between radio and MIR emission, which is true for BZBs. However, the correlations between radio and MIR flux/lumniosity of BZQs are much weaker than those of SFGs and BZBs. This makes the jet origin of their MIR emission not firm. In \textit{WISE} color-color diagram, BZQs also show similar colors with other AGNs, including quasars and Seyfert galaxies~\citep{2011ApJ...735..112J, 2011ApJ...740L..48M, 2022A&A...659A..32P}. If the MIR emission from torus is comparable with that from jet, the mixed MIR emission components can weaken the radio/MIR correlation of BZQs. Then the AGN-like MIR color and weak radio/MIR correlation can be explained naturally. SSRQs and RGs in the SED library locate similar region with BZQs on the color-flux ratio diagram (Figure~\ref{fig:ratio}), which also supports that the MIR emission of BZQs should be dominated by torus.

The MIR color of NLS1s has been explored by several authors.~\citet{2015MNRAS.451.1795C} considered the contributions from jet, torus and host galaxy for a sample of RL NLS1s and concluded that pure jet contribution cannot explain the observed \textit{WISE} colors. SF component was needed to explain the MIR emission at longer wavelength, especially for the objects with redder SEDs (W3 $-$ W4 $> 2.5$). They also suggested that for several red objects with larger W1 $-$ W2, the MIR emission was a mixture of non-thermal jet radiation and torus emission. Other results showed that NLS1s occupy similar region with blazars in \textit{WISE} color-color diagram~\citep{2018MNRAS.480.1796S}, which was also found in a sample of type 1 Seyfert galaxies~\citep{2022A&A...659A..32P}. Our results confirm that most LoTSS detected NLS1s have similar MIR colors with BZQs. In addition, the main NLS1 sample shows MIR excess compared with SFGs. The MIR luminosity of NLS1s is also higher than that of SFGs\footnote{Detailed comparison of MIR and radio luminosity should consider the effect of stellar mass, which will be explored in a future work.}. These facts indicate that the MIR emission of NLS1s contains additional contributions from AGN activity, whether jet or torus. The rest of NLS1s with smaller W1 $-$ W2 locate inside the region of spiral galaxies in \textit{WISE} color-color diagram~\citep{2010AJ....140.1868W}, which could be dominated by the radiation from their host galaxies~\citep{2018A&A...619A..69J}.

The radio emission of the main NLS1 sample is ambiguous. One possibility is that it originates from SF activity, like SFGs. There are several arguments supporting this scenario. First one is the low radio luminosity of them ($\lesssim 10^{32} erg s^{-1} Hz^{-1}$ with the mean value of $4.37\times10^{30} erg s^{-1} Hz^{-1}$). It is suggested that SF process could dominate radio emission when the radio luminosity is lower than about $5\times10^{30} erg s^{-1} Hz^{-1}$ (\citealt{2016A&ARv..24...13P}, converting from 1.4 GHz luminosity into 144 MHz luminosity with a power-law index 0.7). However, low power jet structures at pc- or kpc-scale have been detected in RQ NLS1s and low luminosity AGNs~\citep{2015ApJ...798L..30D, 2020A&A...634A.108S, 2021MNRAS.508.1305Y, 2022A&A...662A..20V}. RGs with radio luminosities lower than this luminosity criterion also exist in our work (Figure~\ref{fig:lum}).

Another result implying the SF origin of the radio emission for the main NLS1 sample is their indistinguishable distribution with SFGs in the luminosity-luminosity diagram (Figure~\ref{fig:lum}), although they show MIR excess when compared with SFGs. Other RQ AGNs were also found to follow the radio/IR correlation of SFGs, and their radio emission has been examined as a tracer of SFR~\citep{2015MNRAS.453.1079B}. However, there are results showing that the radio and IR emission of sources following or slightly deviating from radio/IR correlation cannot simply be treated as SF origin or not, respectively.~\citet{2016MNRAS.460.1588W} explored the radio/FIR correlation and fundamental plane for a sample of hard X-ray selected AGNs and suggested that their radio emission was related to AGN activity rather than SF process, although those AGNs followed the radio/FIR correlation. For their sample, the FIR spectral index, and the $q$ parameter (defined as the ratio between FIR and radio flux density/luminosity) of AGNs were slightly different from those of SFGs.~\citet{2013ApJ...778...94R} explored the radio/MIR properties for a sample of type 2 Seyfert galaxies and found MIR excess for them in $L_{1.4 GHz}$-$L_{12 \mu m}$ and $L_{1.4 GHz}$-$L_{22 \mu m}$ diagrams. They compared these Seyfert galaxies with a sample of SFGs with similar stellar mass and redshift, and found SFGs and Seyfert galaxies were generally overlapped in $L_{1.4 GHz}$-$L_{22 \mu m}$ diagram. In $L_{1.4 GHz}$-$L_{12 \mu m}$ diagram, SFGs showed slightly stronger MIR excess. However, after exploring their properties of 4000 $\AA$ break and specific radio luminosity, they suggested that the radio and MIR emission of Seyfert galaxies was dominated by SF process.

Based on the above arguments, additional information apart from radio/IR correlation and radio luminosity is needed to clarify the exact origin of radio emission for the main NLS1 sample. The strong correlation between radio and MIR emission suggests that the radio and MIR emission of these NLS1s have common origin. As mentioned above, the MIR emission of NLS1s contains additional contributions from AGN activity, which implies a similar AGN origin of their radio emission. In Figure~\ref{fig:ratio}, the main NLS1 sample shows similar, even larger flux ratio of radio to MIR emission than that of Mrk 231, a RQ quasar with weak jet~\citep{2021MNRAS.504.3823W, 2021MNRAS.507.2550S}.~\citet{2021MNRAS.507.2550S} suggested that the AGN related wind of Mrk 231 could exceed SF related one. Mrk 231 is also classified as an ultraluminous infrared galaxy with extreme starburst~\citep{1999ApJ...519..185T}. Thus it can be expected that jet, AGN wind and starburst can also coexist in NLS1s.~\citet{2022A&A...658A..12J} investigated the origin of extended radio emission of NLS1s through radio spectral index map. They found that AGN or SF activity, or combination of both can dominate radio emission of individual NLS1.~\citet{2020Univ....6...45F} found weak correlation between radio emission and broad emission line of LoTSS detected NLS1s, which suggested AGN related radio emission should exist in NLS1s. All these results indicate that both radio and MIR emission of the main NLS1 sample could contain substantial contributions from AGN activities, including jet or central accretion process. Nevertheless, the emission from SF process cannot be rejected neither. To clarify which component is dominant in NLS1s, detailed explorations of the fundamental plane, specific radio/MIR luminosity, or emission line properties for them can be helpful.

A small fraction of NLS1s show similar radio excess with blazars and RGs, which indicates that these NLS1s can also host jets as blazars and RGs. Thus the radio excess deviating from radio/MIR correlation of SFGs can be used to select NLS1s whose radio emission is most likely produced by jet activity. As the radio/MIR correlations of long wavelength are generally better than those of short wavelength for SFGs, and considering the sample size, we choose the correlation of SFGs between 144 MHz radio luminosity and $12 \mu m$ MIR luminosity to select jetted AGN candidates. The 0.5 dex line above this correlation (the upper blue dotted line in Figure~\ref{fig:lum_fit}) is taken as the dividing criterion. Figure~\ref{fig:lum_fit} shows the 3-$\sigma$ outliers of NLS1s (open circles) in $L_{144 MHz}$-$L_{12 \mu m}$ diagram. Except for one object with L$_{12 \mu m} > L_{144 MHz}$, the other 66 objects all exceed the dividing criterion. Thus all these 66 objects are considered as jetted NLS1s candidates. Table~\ref{tab:nls1} lists the detailed information of them. We also note that these sources generally have $L_{144 MHz}$ larger than $5\times10^{30} erg s^{-1} Hz^{-1}$ with the lowest radio luminosity $L_{144 MHz} = 4.90\times10^{30} erg s^{-1} Hz^{-1}$. Among them, there are seven flat spectrum RL NLS1s (ILTJ084958.00+510828.9, ILTJ095317.10+283601.5, ILTJ113824.51+365327.5, ILTJ114654.21+323651.8, ILTJ142114.06+282453.0, ILTJ154817.92+351127.8, and ILTJ163402.03+480940.8) in~\citet{2015A&A...575A..13F} and two objects (ILTJ113824.51+365327.5 and ILTJ130522.77+511640.1) in~\citet{2022A&A...658A..12J}. The radio emission of both objects is confirmed to be AGN related.

\begin{figure*}[h!]
\centering
\includegraphics[angle=0,scale=0.8]{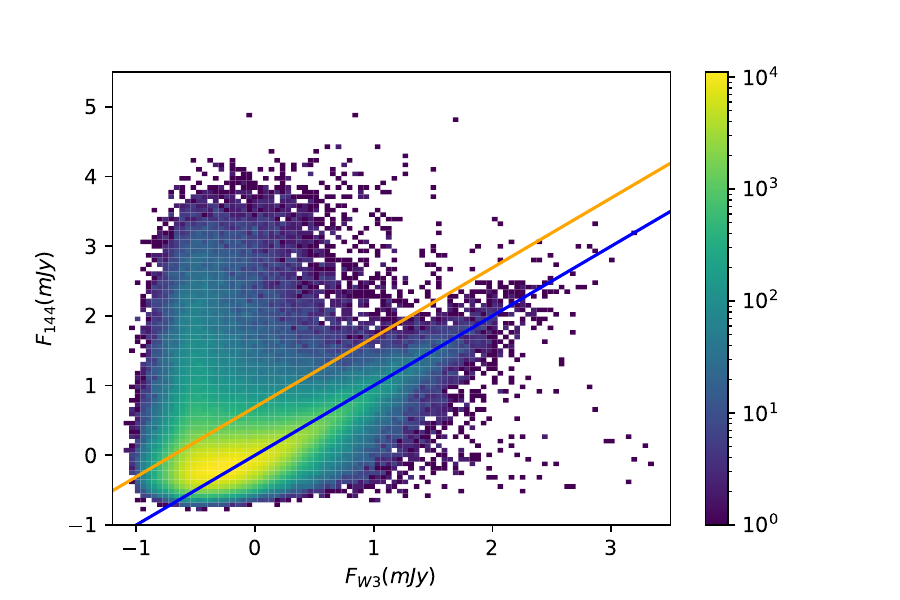}
\caption{The 2D histogram between 144 MHz radio flux density and 12 $\mu m$ MIR flux density for the LoTSS-AllWISE sample. The blue solid line shows the one to one relation, while the orange solid line labels the criterion to distinguish AGNs from SFGs suggested by~\citet{2021ApJ...910...64K} (see the context for details). \label{fig:whole}}
\end{figure*}

\citet{2023A&A...678A.151H} explored the MIR properties for their optically identified LoTSS DR2 catalogue, and found that two distinct branches exhibit on the plot of 144 MHz radio flux density versus W2 magnitude, which trace SF and jet processes, respectively. In Figure~\ref{fig:whole}, we plot the 144 MHz radio flux density and W3 flux density for LoTSS-AllWISE sample in our work. The two branches are also obvious. The SF branch is around the line that 144 MHz radio flux density equals to 12 $\mu m$ MIR flux density (blue solid line in Figure~\ref{fig:whole}). The line that 1.4 GHz radio flux density equals to 12 $\mu m$ MIR flux density was suggested to distinguish radio AGNs from SFGs~\citep{2021ApJ...910...64K}. We convert the 1.4 GHz radio flux density into 144 MHz with a spectral index $\alpha_{radio} = 0.7$, and plot this criterion as an orange line in the lower left panel of Figure~\ref{fig:flux} and Figure~\ref{fig:whole}. In Figure~\ref{fig:flux}, most SFGs and jetted AGNs (RGs and blazars) locate below and above this criterion, respectively. Most NLS1s with radio excess also locate above this criterion. In Figure~\ref{fig:whole}, this criterion seems also valid to separate jetted AGNs from SFGs, although there are plenty of objects distributing around the overlapping region.

\section{Conclusion}\label{sec:summary}
In this paper, we systematically compare the radio/MIR correlations and MIR colors among SFGs, RGs, BZBs, BZQs, as well as NLS1s with both LoTSS and \textit{WISE} observations. BZBs, BZQs and RGs show obvious radio excess in radio-MIR diagram, which is consistent with the jet origin of their radio emission. BZBs show the most significant correlation between radio and MIR emission among these five classes of sources, while BZQs show much weaker correlations between low-frequency radio and MIR emission. Weak correlations for RGs are only found at W1 and W2 bands. The MIR colors (W1 $-$ W2 and W2 $-$ W3) of BZBs and BZQs are redder than those of RGs. These differences suggest different origins for their MIR emission. MIR emission of RGs is dominated by old stellar populations in their host galaxies. For BZBs, it is dominated by the non-thermal emission from jet. Mixed jet and torus components can explain the weak correlation of BZQs. A small fraction of NLS1s show similar radio excess with RGs and blazars, which suggests that their radio emission is also dominated by jet. We obtain a sample of jetted NLS1 candidates by identifying NLS1s with radio excess. The tight radio/MIR correlation is confirmed for SFGs. The majority of NLS1s also show significant correlations between low-frequency radio and MIR emission. But their emission origins should be different. The radio and MIR emission of SFGs is believed to be produced by SF activities. The redder W1 $-$ W2 color and the MIR excess of NLS1s in radio-MIR diagram indicate substantial AGN contributions to their MIR emission. Considering the tight correlation between radio and MIR emission, the radio emission of NLS1s can also be a combination of AGN and SF related components. In addition, the flux-flux diagram between 144 MHz radio flux density and $12 \mu m$ MIR flux density is confirmed to be valid to distinguish radio AGNs from SFGs.

\acknowledgments
We thank the anonymous referee for the constructive comments that improve
our manuscript greatly. This research is funded by National Natural Science Foundation of China (NSFC; grant number 12003014).

This publication makes use of data products from LOFAR. LOFAR data products were provided by the LOFAR Surveys Key Science project (LSKSP; https://lofar-surveys.org/) and were derived from observations with the International LOFAR Telescope (ILT). LOFAR~\citep{2013A&A...556A...2V} is the Low Frequency Array designed and constructed by ASTRON. It has observing, data processing, and data storage facilities in several countries, which are owned by various parties (each with their own funding sources), and which are collectively operated by the ILT foundation under a joint scientific policy. The efforts of the LSKSP have benefited from funding from the European Research Council, NOVA, NWO, CNRS-INSU, the SURF Co-operative, the UK Science and Technology Funding Council and the J¨¹lich Supercomputing Centre.
This publication makes use of data products from the Wide-field Infrared Survey Explorer, which is a joint project of the University of California, Los Angeles, and the Jet Propulsion Laboratory/California Institute of Technology, funded by the National Aeronautics and Space Administration.
This research has made use of the VizieR catalogue access tool, CDS,
Strasbourg, France (DOI : 10.26093/cds/vizier). The original description
of the VizieR service was published in~\citet{2000A&AS..143...23O}.

%

\vspace{5mm}

\facilities{LOFAR (LoTSS DR2), \textit{WISE} (AllWISE)}


\software{Astropy~\citep{2013A&A...558A..33A}, TOPCAT~\citep{2005ASPC..347...29T}, linmix~\citep{2007ApJ...657..116K}, speclite~\citep{2023zndo...7734526K}, Scipy, Pandas, Scikit-learn, Matplotlib}


\bibliography{bib}
\bibliographystyle{aasjournal}

\begin{longrotatetable}
\begin{deluxetable*}{cccccccccccc}
\tablecaption{The Properties of Jetted NLS1 Candidates \label{tab:nls1}}
\tabletypesize{\tiny}

\tablehead{
\colhead{LoTSS Name} & \colhead{RA} &
\colhead{DEC} & \colhead{z} &
\colhead{$F_{144 MHz}$} & \colhead{Error} &
\colhead{$\log L_{144 MHz}$} & \colhead{Error} &
\colhead{W3 mag} & \colhead{Error} & \colhead{$\log L_{W3}$}  & \colhead{Error}\\
\colhead{} & \colhead{} & \colhead{} & \colhead{} & \colhead{(mJy)} &
\colhead{(mJy)} & \colhead{($erg s^{-1} Hz^{-1}$)} & \colhead{($erg s^{-1} Hz^{-1}$)} &
\colhead{(mag)} & \colhead{(mag)} & \colhead{($erg s^{-1} Hz^{-1}$)} & \colhead{($erg s^{-1} Hz^{-1}$)}
}
\startdata
ILTJ075114.82+251632.6 & 117.81176 & 25.27573 & 0.3364 & 17.84 & 0.26 & 31.79 & 6.4e-03 & 10.24 & 0.08 & 30.98 & 0.03 \\
ILTJ083314.42+270331.7 & 128.31008 & 27.05882 & 0.0660 & 47.80 & 1.28 & 30.69 & 1.2e-02 & 9.73 & 0.05 & 29.63 & 0.02 \\
ILTJ083454.85+553420.5 & 128.72854 & 55.57237 & 0.2415 & 9273.84 & 7.56 & 34.18 & 3.5e-04 & 10.52 & 0.08 & 30.54 & 0.03 \\
ILTJ084051.57+522202.7 & 130.21486 & 52.36743 & 0.7210 & 19.39 & 0.27 & 32.58 & 6.1e-03 & 11.54 & 0.19 & 31.25 & 0.08 \\
ILTJ084958.00+510828.9 & 132.49167 & 51.14138 & 0.5840 & 193.68 & 1.20 & 33.37 & 2.7e-03 & 10.04 & 0.05 & 31.63 & 0.02 \\
ILTJ085001.18+462600.7 & 132.50490 & 46.43354 & 0.5237 & 69.25 & 0.58 & 32.82 & 3.6e-03 & 11.50 & 0.19 & 30.93 & 0.08 \\
ILTJ085022.21+450101.3 & 132.59254 & 45.01705 & 0.5421 & 73.65 & 0.42 & 32.88 & 2.5e-03 & 10.48 & 0.07 & 31.38 & 0.03 \\
ILTJ090409.57+581527.1 & 136.03988 & 58.25753 & 0.1461 & 23.96 & 0.35 & 31.12 & 6.4e-03 & 10.95 & 0.12 & 29.88 & 0.05 \\
ILTJ091953.31+413911.1 & 139.97212 & 41.65309 & 0.7737 & 130.69 & 1.03 & 33.48 & 3.4e-03 & 11.23 & 0.15 & 31.45 & 0.06 \\
ILTJ092632.22+543847.3 & 141.63424 & 54.64649 & 0.4125 & 122.62 & 0.82 & 32.83 & 2.9e-03 & 10.72 & 0.09 & 31.00 & 0.03 \\
ILTJ093241.15+530633.5 & 143.17145 & 53.10931 & 0.5970 & 578.11 & 0.91 & 33.87 & 6.8e-04 & 10.75 & 0.08 & 31.37 & 0.03 \\
ILTJ093712.38+500852.7 & 144.30157 & 50.14798 & 0.2755 & 70.65 & 0.62 & 32.19 & 3.8e-03 & 8.95 & 0.03 & 31.30 & 0.01 \\
ILTJ094502.93+553036.8 & 146.26219 & 55.51024 & 0.6448 & 8.48 & 0.14 & 32.11 & 7.2e-03 & 11.03 & 0.10 & 31.34 & 0.04 \\
ILTJ095317.10+283601.5 & 148.32127 & 28.60043 & 0.6577 & 143.31 & 1.99 & 33.36 & 6.0e-03 & 11.34 & 0.18 & 31.24 & 0.07 \\
ILTJ095446.68+585734.1 & 148.69449 & 58.95947 & 0.7819 & 27.04 & 0.44 & 32.81 & 7.1e-03 & 12.04 & 0.20 & 31.14 & 0.08 \\
ILTJ095820.99+322402.6 & 149.58745 & 32.40073 & 0.5306 & 5213.40 & 3.98 & 34.71 & 3.3e-04 & 9.21 & 0.04 & 31.86 & 0.01 \\
ILTJ095909.50+460014.1 & 149.78957 & 46.00393 & 0.3989 & 15.47 & 0.33 & 31.89 & 9.4e-03 & 11.18 & 0.14 & 30.78 & 0.06 \\
ILTJ100410.95+523024.6 & 151.04562 & 52.50685 & 0.2987 & 28.05 & 0.34 & 31.87 & 5.3e-03 & 12.10 & 0.31 & 30.12 & 0.12 \\
ILTJ100441.44+403647.9 & 151.17269 & 40.61332 & 0.4999 & 126.98 & 0.74 & 33.03 & 2.5e-03 & 11.81 & 0.24 & 30.76 & 0.10 \\
ILTJ101422.57+400717.4 & 153.59403 & 40.12150 & 0.6727 & 9.20 & 0.19 & 32.19 & 8.9e-03 & 11.06 & 0.14 & 31.37 & 0.05 \\
ILTJ101840.61+282109.7 & 154.66920 & 28.35269 & 0.3841 & 21.03 & 1.26 & 31.99 & 2.6e-02 & 11.99 & 0.32 & 30.42 & 0.13 \\
ILTJ103407.73+401510.2 & 158.53221 & 40.25286 & 0.3733 & 97.68 & 0.52 & 32.63 & 2.3e-03 & 10.51 & 0.08 & 30.98 & 0.03 \\
ILTJ104136.96+391114.3 & 160.40401 & 39.18732 & 0.7150 & 5.13 & 0.12 & 32.00 & 1.0e-02 & 12.62 & 0.51 & 30.81 & 0.20 \\
ILTJ110249.81+525012.5 & 165.70753 & 52.83682 & 0.6898 & 41.51 & 0.22 & 32.87 & 2.3e-03 & 11.78 & 0.22 & 31.11 & 0.09 \\
ILTJ111005.13+453213.4 & 167.52138 & 45.53706 & 0.6130 & 8.85 & 0.29 & 32.08 & 1.4e-02 & 11.96 & 0.31 & 30.91 & 0.12 \\
ILTJ111106.37+520814.7 & 167.77655 & 52.13742 & 0.4950 & 9.56 & 0.25 & 31.90 & 1.1e-02 & 11.01 & 0.09 & 31.07 & 0.04 \\
ILTJ111450.07+502917.3 & 168.70861 & 50.48816 & 0.4518 & 29.72 & 0.71 & 32.30 & 1.0e-02 & 11.34 & 0.14 & 30.84 & 0.06 \\
ILTJ111818.01+385057.0 & 169.57503 & 38.84918 & 0.6202 & 4.94 & 0.19 & 31.84 & 1.7e-02 & 12.10 & 0.34 & 30.87 & 0.13 \\
ILTJ113646.02+314108.1 & 174.19176 & 31.68561 & 0.2325 & 8.69 & 0.33 & 31.12 & 1.6e-02 & 11.20 & 0.14 & 30.23 & 0.06 \\
ILTJ113824.51+365327.5 & 174.60213 & 36.89099 & 0.3563 & 52.26 & 0.93 & 32.31 & 7.7e-03 & 10.53 & 0.08 & 30.92 & 0.03 \\
ILTJ114628.47+651353.7 & 176.61861 & 65.23159 & 0.7821 & 17.59 & 0.36 & 32.62 & 8.8e-03 & 11.93 & 0.23 & 31.19 & 0.09 \\
ILTJ114654.21+323651.8 & 176.72586 & 32.61440 & 0.4649 & 36.35 & 0.51 & 32.42 & 6.0e-03 & 10.51 & 0.08 & 31.20 & 0.03 \\
ILTJ114748.90+331605.9 & 176.95376 & 33.26831 & 0.5803 & 15.63 & 0.29 & 32.27 & 8.0e-03 & 10.75 & 0.12 & 31.34 & 0.05 \\
ILTJ115255.22+610604.8 & 178.23009 & 61.10134 & 0.2918 & 11.15 & 0.11 & 31.45 & 4.5e-03 & 10.60 & 0.07 & 30.69 & 0.03 \\
ILTJ115338.52+643750.8 & 178.41049 & 64.63080 & 0.4441 & 21.70 & 0.12 & 32.15 & 2.4e-03 & 10.98 & 0.11 & 30.97 & 0.04 \\
ILTJ115632.50+555926.0 & 179.13541 & 55.99057 & 0.2900 & 14.80 & 0.29 & 31.56 & 8.4e-03 & 11.23 & 0.12 & 30.44 & 0.05 \\
ILTJ115700.71+324458.2 & 179.25296 & 32.74951 & 0.4862 & 976.15 & 3.51 & 33.89 & 1.6e-03 & 10.43 & 0.08 & 31.28 & 0.03 \\
ILTJ120027.34+601950.4 & 180.11392 & 60.33068 & 0.6250 & 25.60 & 0.13 & 32.56 & 2.2e-03 & 10.58 & 0.08 & 31.49 & 0.03 \\
ILTJ120910.62+561109.5 & 182.29426 & 56.18597 & 0.4532 & 8.64 & 0.16 & 31.77 & 7.8e-03 & 11.48 & 0.15 & 30.79 & 0.06 \\
ILTJ123451.47+353358.4 & 188.71445 & 35.56623 & 0.6131 & 54.73 & 0.57 & 32.87 & 4.5e-03 & 11.52 & 0.19 & 31.09 & 0.07 \\
ILTJ125316.18+260424.8 & 193.31740 & 26.07356 & 0.4087 & 5.65 & 0.24 & 31.48 & 1.9e-02 & 11.29 & 0.14 & 30.76 & 0.06 \\
ILTJ125623.01+350131.1 & 194.09587 & 35.02533 & 0.4015 & 16.24 & 0.59 & 31.92 & 1.6e-02 & 10.63 & 0.08 & 31.01 & 0.03 \\
ILTJ130522.77+511640.1 & 196.34486 & 51.27782 & 0.7850 & 276.36 & 2.57 & 33.82 & 4.0e-03 & 8.91 & 0.03 & 32.39 & 0.01 \\
ILTJ132810.22+455241.5 & 202.04256 & 45.87821 & 0.4391 & 15.54 & 0.46 & 31.99 & 1.3e-02 & 11.34 & 0.13 & 30.81 & 0.05 \\
ILTJ133508.35+671734.4 & 203.78478 & 67.29289 & 0.4946 & 7.07 & 0.12 & 31.77 & 7.3e-03 & 13.20 & 0.49 & 30.19 & 0.19 \\
ILTJ134540.04+280123.2 & 206.41683 & 28.02312 & 0.1663 & 40.03 & 1.21 & 31.46 & 1.3e-02 & 10.62 & 0.08 & 30.13 & 0.03 \\
ILTJ140222.73+310302.2 & 210.59470 & 31.05062 & 0.5794 & 4.72 & 0.17 & 31.75 & 1.5e-02 & 12.52 & 0.31 & 30.63 & 0.12 \\
ILTJ141324.23+493425.1 & 213.35097 & 49.57365 & 0.3708 & 3.59 & 0.13 & 31.19 & 1.5e-02 & 11.85 & 0.16 & 30.43 & 0.06 \\
ILTJ141403.18+352311.6 & 213.51324 & 35.38656 & 0.0574 & 82.49 & 0.64 & 30.81 & 3.4e-03 & 8.35 & 0.02 & 30.06 & 0.01 \\
ILTJ141557.26+495333.8 & 213.98859 & 49.89275 & 0.1854 & 274.94 & 0.99 & 32.40 & 1.6e-03 & 9.12 & 0.03 & 30.84 & 0.01 \\
ILTJ142106.04+385522.8 & 215.27519 & 38.92302 & 0.4890 & 169.69 & 1.84 & 33.14 & 4.7e-03 & 9.64 & 0.03 & 31.60 & 0.01 \\
ILTJ142114.06+282453.0 & 215.30860 & 28.41474 & 0.5380 & 54.00 & 1.07 & 32.74 & 8.6e-03 & 9.61 & 0.04 & 31.71 & 0.02 \\
ILTJ143244.92+301435.4 & 218.18718 & 30.24318 & 0.3546 & 335.49 & 1.82 & 33.11 & 2.4e-03 & 10.38 & 0.05 & 30.98 & 0.02 \\
ILTJ143509.52+313148.0 & 218.78965 & 31.53002 & 0.5019 & 236.68 & 1.20 & 33.31 & 2.2e-03 & 9.97 & 0.04 & 31.50 & 0.02 \\
ILTJ143627.08+524839.4 & 219.11284 & 52.81095 & 0.4151 & 13.81 & 0.59 & 31.88 & 1.9e-02 & 10.87 & 0.07 & 30.94 & 0.03 \\
ILTJ144043.38+613008.9 & 220.18074 & 61.50250 & 0.4423 & 13.49 & 0.11 & 31.94 & 3.6e-03 & 11.33 & 0.10 & 30.82 & 0.04 \\
ILTJ152139.20+285627.1 & 230.41333 & 28.94087 & 0.4890 & 33.15 & 0.99 & 32.43 & 1.3e-02 & 11.55 & 0.13 & 30.84 & 0.05 \\
ILTJ154510.98+345246.6 & 236.29577 & 34.87962 & 0.5160 & 96.83 & 1.40 & 32.95 & 6.3e-03 & 8.83 & 0.02 & 31.98 & 0.01 \\
ILTJ154817.92+351127.8 & 237.07468 & 35.19107 & 0.4786 & 337.14 & 7.31 & 33.41 & 9.4e-03 & 9.88 & 0.04 & 31.48 & 0.01 \\
ILTJ155644.15+531357.4 & 239.18397 & 53.23263 & 0.5628 & 31.06 & 0.31 & 32.54 & 4.3e-03 & 9.88 & 0.04 & 31.65 & 0.01 \\
ILTJ161651.94+601922.3 & 244.21640 & 60.32286 & 0.4233 & 21.27 & 0.57 & 32.09 & 1.2e-02 & 10.50 & 0.04 & 31.11 & 0.02 \\
ILTJ162458.37+423107.3 & 246.24321 & 42.51870 & 0.6635 & 16.02 & 0.35 & 32.42 & 9.5e-03 & 10.67 & 0.08 & 31.51 & 0.03 \\
ILTJ163402.03+480940.8 & 248.50844 & 48.16135 & 0.4947 & 35.26 & 1.02 & 32.47 & 1.3e-02 & 10.71 & 0.05 & 31.19 & 0.02 \\
ILTJ163559.36+304032.6 & 248.99735 & 30.67574 & 0.5785 & 56.98 & 0.65 & 32.83 & 5.0e-03 & 9.51 & 0.04 & 31.83 & 0.02 \\
ILTJ165247.37+411548.6 & 253.19736 & 41.26350 & 0.5093 & 217.23 & 2.01 & 33.28 & 4.0e-03 & 12.42 & 0.33 & 30.53 & 0.13 \\
ILTJ165605.51+444050.4 & 254.02297 & 44.68069 & 0.2261 & 52.08 & 0.52 & 31.87 & 4.4e-03 & 12.14 & 0.23 & 29.82 & 0.09 \\
\enddata
\end{deluxetable*}
\end{longrotatetable}



\end{document}